\documentclass[prl,twocolumn,showpacs,superscriptaddress]{revtex4}
\usepackage{graphicx}
\usepackage{amsfonts}
\usepackage{amsmath}
\usepackage{amssymb}
\usepackage{bm}

\usepackage{amsmath}
\usepackage{color}
\usepackage{textcomp}
\usepackage[colorlinks=true, letterpaper=true, pdfstartview=FitV, linkcolor=blue, citecolor=blue, urlcolor=blue]{hyperref}

\begin{document}
	
\title{Probing the direct factor for superconductivity in FeSe-Based Superconductors by Raman Scattering}

\author{Anmin Zhang}

\affiliation{School of Physical Science and Technology, Lanzhou University, Lanzhou 730000, China}
\author{Xiaoli Ma}
\author{Yimeng Wang}
\author{Shanshan Sun}
\affiliation{Department of Physics, Renmin University of China, Beijing 100872, China}
\author{Bin Lei}
\affiliation{Hefei National Laboratory for Physical Science at Microscale and Department of Physics, University of Science and Technology of China, Hefei, Anhui 230026, China}
\author{Hechang Lei}
\affiliation{Department of Physics, Renmin University of China, Beijing 100872, China}
\author{Xianhui Chen}
\affiliation{Hefei National Laboratory for Physical Science at Microscale and Department of Physics, University of Science and Technology of China, Hefei, Anhui 230026, China}
\author{Xiaoqun Wang}
\affiliation{Department of Physics and Astronomy, Shanghai Jiao Tong University, Shanghai 200240, China}
\author{Changfeng Chen}
\affiliation{Department of Physics and Astronomy, University of Nevada, Las Vegas, Nevada 89154, USA}
\author{Qingming Zhang}
\email{qmzhang@ruc.edu.cn}
\affiliation{School of Physical Science and Technology, Lanzhou University, Lanzhou 730000, China}
\affiliation{Beijing National Laboratory for Condensed Matter Physics, Institute of Physics, Chinese Academy of Sciences, Beijing 100190, China}

\date{\today}

\begin{abstract}
The FeSe-based superconductors exhibit a wide range of critical temperature $T_c$ under a variety of material and physical conditions, but extensive studies to date have yet to produce a consensus view on the underlying mechanism. Here we report on a systematic Raman scattering work on intercalated FeSe superconductors Li$_x$(NH$_3$)$_y$Fe$_2$Se$_2$ and (Li,Fe)OHFeSe compared to pristine FeSe. All three crystals show an anomalous power-law temperature dependence of phonon linewidths, deviating from the standard anharmonic behavior. This intriguing phenomenon is attributed to electron-phonon coupling effects enhanced by electron correlation, as evidenced by the evolution of the $A_{1g}$ Raman mode. Meanwhile, an analysis of the $B_{1g}$ mode, which probes the out-of-plane vibration of Fe, reveals a lack of influence by previously suggested structural parameters, and instead indicates a crucial role of the joint density of states in determining $T_c$. These findings identify carrier doping as the direct factor driving and modulating superconductivity in FeSe-based compounds.

\end{abstract}

\pacs{74.70-b, 74.62.Bf, 63.20.Kr, 78.30.-j}

\maketitle
Iron-based superconductors are a remarkable class of materials that offer a distinct paradigm to elucidate novel physics of high-temperature
superconductivity \cite{Kamihara}. A prominent structural feature of FeSe or FeAs layers commonly appears in this class of materials \cite{Paglione}. Pristine FeSe in the PbO structure has a superconducting critical temperature $ T_{c} $ $ \sim $8 K at ambient pressure \cite{Hsu}. Material and physical conditions like doping, pressure and liquid-gating can substantially enhance $ T_{c} $, up to 14 K by isovalent substitution of Se \cite{Fang}, 37 K by pressure \cite{Medvedev}, and 48 K by liquid-gating \cite{Lei}. Some FeSe-derived superconductors also can host high $ T_{c}s $, such as K$_{x}$Fe$_{2-y}$Se$_{2}$ with an Fe-vacancy order \cite{Guo}, A$_{x}$Fe$_2$Se$_2$ (A = Li, Na, Ba, Sr, Ca, Yb, and Eu) \cite{Ying}, and the intercalated systems Li$_{x}$(NH$_{2}$)$_{y}$(NH$_{3}$)$_{1-y}$Fe$_{2}$Se$_{2}$ \cite{Burrard-Lucas}, Li$_x$(NH$_3$)$_y$Fe$_2$Se$_2$ \cite{Scheidt} and (Li,Fe)OHFeSe \cite{Lu,Krzton-Maziopa,Zhang}. These latter compounds contain alkali-metal ions, ammonia, or organic molecules intercalated between adjacent FeSe layers and exhibit $ T_{c} $ values of about 44 K. It has been reported that monolayer FeSe on a SrTiO$_{3}$ substrate reached a very high $T_{c}$ above 100 K \cite{Wang,Ge}. The FeSe-based superconductors have $T_c$s comparable to those of FeAs-based superconductors, but unlike the latter have no insulating and magnetically ordered parent phase \cite{Hsu}. Fermi surface nesting disappears in some FeSe-based superconductors due to vanishing hole pockets \cite{Liu}, although both systems display a nematic phase \cite{Nakayama,Fernandes}. These property distinctions are considered indications of different pairing mechanisms in these two series of iron-based superconductors \cite{Mazin,Kontani}.

The magnetism, nematicity and superconductivity in FeSe-based superconductors are intimately connected to the crystal structure \cite{Medvedev,Bendele,Bendele1,Kn,Kaluarchchi,Terashima,Sun,Sun1,Massat}.
Pressure can simultaneously change $T_{c}$, magnetism and nematicity by driving changes in lattice parameters \cite{Medvedev,Bendele,Bendele1,Kn,Kaluarchchi,Terashima,Sun,Sun1,Massat}. Theoretical studies have proposed that anion heights or Se-Fe-Se bond angles may be key factors in determining Tc in iron-based superconductors \cite{Kuroki,Moon}. This proposal seems valid in most of FeAs-based superconductors, where $T_{c}$ peaks when the FeAs$_{4}$ unit comes close to a regular tetrahedron shape or when the anion height is close to the optimal value (1.38 \AA{}) \cite{Mizuguchi,Okabe}.  This idea was also applied to pressurized and Te/S co-doped FeSe \cite{Mizuguchi,Okabe}, but no conclusive evidence has been established in FeSe-based superconductors. Instead, various structural parameters have been proposed as driving mechanisms, including the separation between neighboring FeSe layers \cite{Zhang}, distortion of the FeSe$_{4}$ unit, and the deviation of anion height from the optimum value (1.38 \AA{}) \cite{Lu1,Lu2}. It was even reported that $T_{c}$ shows a V-shape dependence on anion height \cite{Lu2}. Besides, carrier doping \cite{Lei} and/or the resulting shape of Fermi surfaces \cite{Liu} can also change $T_c$ significantly. It remains unclear which one among these factors, is the direct control factor in determining $T_c$. These inconsistencies hinder a general understanding of FeSe-based superconductors. It is therefore a pressing task to devise a reliable approach capable of assessing the
microscopic mechanism for superconductivity in these novel materials.

In this work, we present a systematic temperature-dependent Raman scattering study of Li$_x$(NH$_3$)$_y$Fe$_2$Se$_2$ and (Li,Fe)OHFeSe single crystals compared to FeSe. Both intercalated compounds exhibit clear $A_{1g}$ and $B_{1g}$ phonon modes showing similar temperature dependence as seen in FeSe, which means that these FeSe-based superconductors contain intact FeSe layers just as in FeSe. Interestingly, the temperature dependence of the linewidths for the two phonon modes in all three crystals do not exhibit the standard anharmonic phonon behavior, but instead follow power laws. This intriguing phenomenon is attributed to electron-phonon coupling (EPC) effects. Among these three systems the frequencies of the $B_{1g}$ mode remain nearly unchanged while those of the $A_{1g}$ mode show a clear variation stemming from the material difference in adjacent intercalated layers. Moreover, the Fano asymmetric lineshapes of the $A_{1g}$ modes in both intercalated compounds suggest appreciable EPC in the systems. The high sensitivity of the $B_{1g}$ mode to the Se height allows a reliable determination that $T_{c}$ shows little dependence to the Se height but has a positive correlation with the joint density of states of occupied and empty states near the Fermi surface. The present findings shed new light on the microscopic mechanism for superconductivity in FeSe-based compounds.

We have grown and characterized FeSe, (Li,Fe)OHFeSe, and Li$_x$(NH$_3$)$_y$Fe$_2$Se$_2$ single crystals following the procedures described elsewhere \cite{Lu,Wang1,Sun2,Lei1}. Magnetization measurements were performed with a Quantum Design magnetic property measurement system (MPMS3) or a Quantum Design physical property measurement system (QD PPMS-14 T). Raman measurements were performed with a Jobin Yvon HR800 single-grating-based micro-Raman system equipped with a volume Bragg grating low-wavenumber suite, a liquid nitrogen-cooled back-illuminated charge-coupled device detector and a 633 nm laser (Melles Griot).  The laser was focused into a spot of $ \sim $ 5 $\mu$m in diameter on sample surface, with a power $ < $ 100$ \mu$W, to avoid overheating. The polarization configuration is denoted by ($ e_{i}, e_{s}$). The notations of \textit{x'} and \textit{y'} were used for the orthorhombic axes which are rotated by 45\textdegree   from the the tetragonal crystallographic axes \textit{a} and \textit{b}. Raman measurements on Li$_x$(NH$_3$)$_y$Fe$_2$Se$_2$ were performed below 200 K to avoid possible sample damage.

\begin{figure}[t]
	\includegraphics[angle=0,width=8cm]{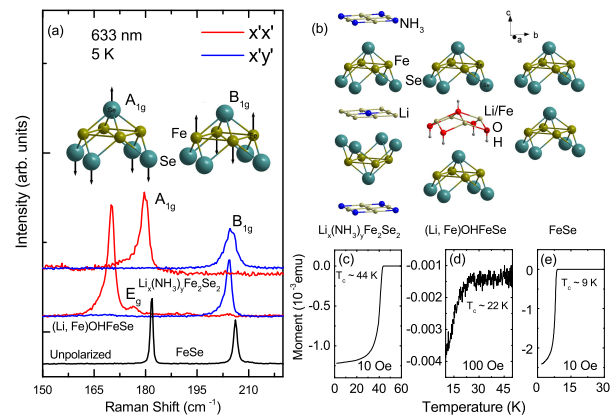}
	\caption{(a) Raman spectra of Li$_x$(NH$_3$)$_y$Fe$_2$Se$_2$, (Li,Fe)OHFeSe and FeSe at low temperatures. The inset shows the atomic displacement patterns of the $A_{1g}$ and $B_{1g}$ modes. (b) The crystal structures and (c-e) diamagnetization measurements for the three superconductors.}
	\label{fig1}
\end{figure}

Li$_x$(NH$_3$)$_y$Fe$_2$Se$_2$, (Li,Fe)OHFeSe and FeSe all contain the same primitive conductive FeSe layer. Li$_x$(NH$_3$)$_y$Fe$_2$Se$_2$ and (Li,Fe)OHFeSe have a Li$_x$(NH$_3$)$_y$ or (Li,Fe)OH layer intercalated between adjacent FeSe layers [Fig. 1(b)], and measured $ T_{c} $ increases from 9 K in FeSe to 22 K in  (Li,Fe)OHFeSe and 44 K in Li$_x$(NH$_3$)$_y$Fe$_2$Se$_2$ [Fig. 1 (c-e)]. With the incident and scattered polarizations lying in the $ab$ plane, symmetry analysis allows two Raman-active modes: $A_{1g}$ (Se) and $B_{1g}$ (Fe). The $A_{1g}$ mode is dominated by the in-phase vibration of Se anions in the same layer, while the $B_{1g}$ mode corresponds to the anti-phase vibration of Fe ions in the same layer \cite{Xia,Gnezdilov}, as illustrated in the inset of Fig. 1(a). According to the Raman tensors, the $A_{1g}$ mode is visible in the \textit{x'x'} channel but vanishes in the \textit{x'y'} channel, while the order is exactly reversed for the $B_{1g}$ mode. This contrasting situation allows convenient and convincing assignment of the phonon modes in FeSe layers via choosing polarization configurations.

We present in Fig. 2 the temperature evolution of Raman spectra for the three crystals. They share similar temperature evolution patterns with increasing temperature, including peak position softening, width broadening and intensity reducing. The spectra of the intercalated compounds are consistent with those of the pristine FeSe in all the key aspects \cite{Gnezdilov}, and are totally different from the spectra of K$_{0.8}$Fe$_{1.6}$Se$_{2}$ with Fe-vacancy ordering \cite{Zhang1}. The close similarities in spectral evolution, mode symmetry and the numbers of observed modes provide strong evidence that the FeSe layers in the intercalated compounds remain intact rather than element-deficient. In contrast to FeSe, however, the intercalated compounds exhibit clear $A_{1g}$ asymmetry that can be attributed to EPC \cite{Klein}. Theoretical studies suggest that EPC in FeSe-based superconductors is pressure-dependent \cite{Mandal} and can be enhanced by antiferromagnetism \cite{Coh}, which have received some experimental support \cite{Gerber}.

\begin{figure}[t]
	\includegraphics[angle=0,width=8cm]{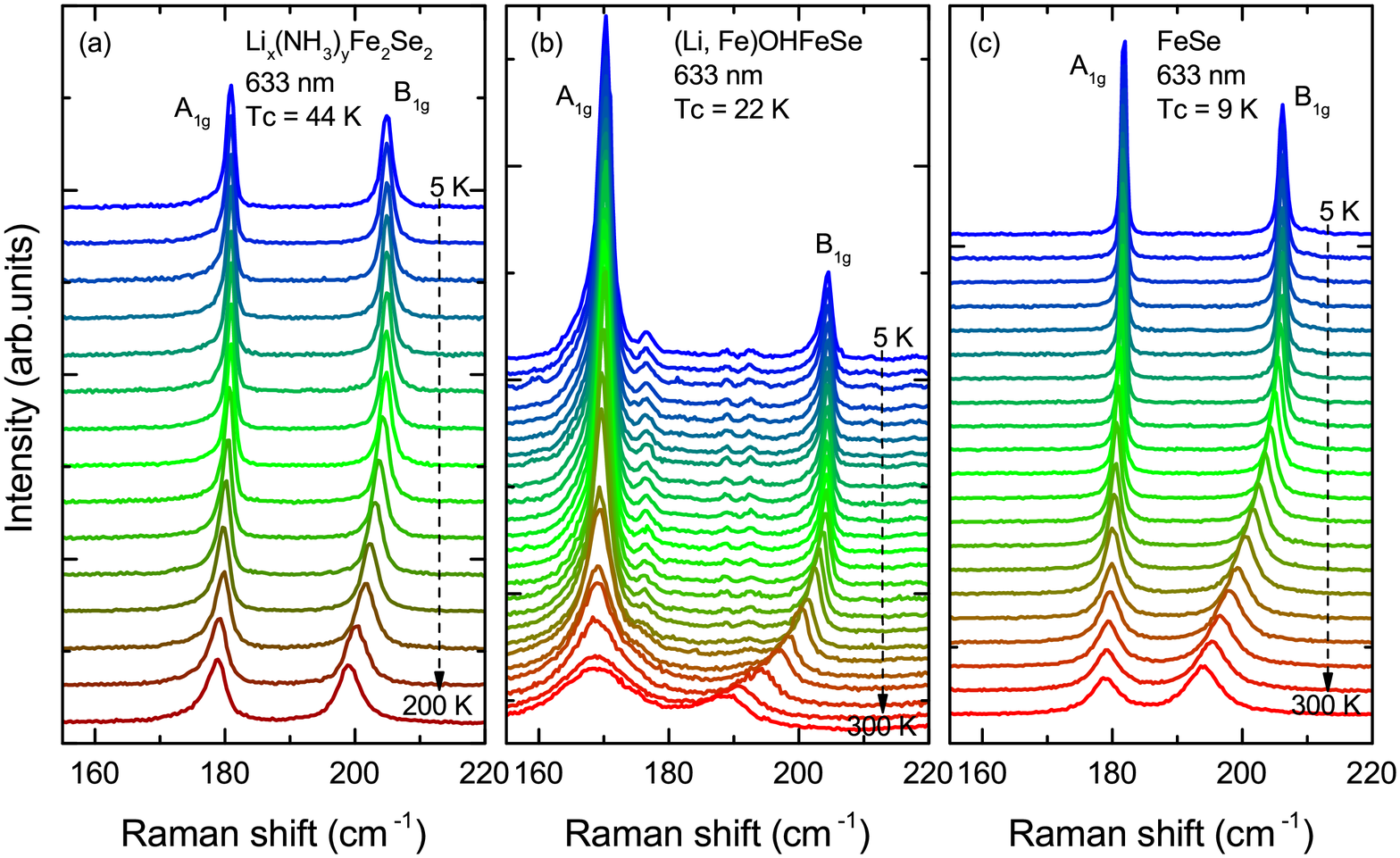}
	\caption{Temperature evolution of Raman spectra of Li$_x$(NH$_3$)$_y$Fe$_2$Se$_2$, (Li,Fe)OHFeSe and FeSe.}
	\label{fig2}
\end{figure}

We have extracted the temperature dependence of frequencies and linewidths of the $A_{1g}$ and $B_{1g}$ modes using Lorentzian and Fano fitting schemes (an example of fitting is shown in Supplemental Material, Fig. S1 \cite{SM}). Results in Fig. 3 show that the $A_{1g}$ frequencies in Li$_x$(NH$_3$)$_y$Fe$_2$Se$_2$ are close to that in FeSe while the frequencies in (Li,Fe)OHFeSe are downshifted by about 10 cm$^{-1}$ [Fig. 3(a)]. Generally speaking, the frequencies of phonon modes are related to crystal structure, lattice parameters and the environment around the vibrating atoms. In our case, the frequencies of $B_{1g}$ mode are similar in the three samples, indicating that the samples have the similar crystal structure and lattice parameters. This difference can be explained by the vibrating pattern of the $A_{1g}$ mode and distinct intercalated layers in the two samples.  The $A_{1g}$ mode represents vibrations of the Se atoms along the \textit{c} axis [see inset of Fig. 1(a)]. Compared to the Fe layers, the Se layers are more susceptible to the influence of adjacent intercalated layers that contain charged OH clusters in (Li,Fe)OHFeSe but electrically neutral NH$_{3}$ in Li$_x$(NH$_3$)$_y$Fe$_2$Se$_2$  [Fig. 1(b)].

By contrast, the $B_{1g}$ mode is dominated by vibrations of the Fe atoms, and is less affected by the intercalated clusters because of the screening by the two Se layers [see the inset of Fig. 1(a)]. Therefore the $B_{1g}$ mode probes more of the features of the interior and intrinsic state of the FeSe layer. The $B_{1g}$ frequencies in the three samples are close to each other, especially in Li$_x$(NH$_3$)$_y$Fe$_2$Se$_2$ and (Li,Fe)OHFeSe, where the frequencies are almost identical at all temperatures. This means the FeSe layers in these samples are in quite similar conditions. The slightly higher $B_{1g}$ frequencies in FeSe can be explained by its larger Se-Fe-Se bond angle and smaller bond distance \cite{Lu}, and according to the Gr\"{u}neisen law, such an effective structural compression generally enhances phonon frequencies (see Table I).

The temperature dependence of the phonon frequencies of the two intercalated compounds [Fig. 3(a), (c)] are well described by a standard anharmonic fitting \cite{Balkanski}, where the frequency and the width are contributed by the three-phonon and four-phonon processes. Surprisingly, however, the linewidths [Fig. 3(b), (d)] show marked deviation from the same fitting scheme, especially for (Li,Fe)OHFeSe, and a power-law fitting yields better agreement with the experimental data. In the high-temperature limit, the two factors in the anharmonic fitting function vary as $T$ and $T^2$, respectively \cite{Balkanski}. But in Figs 3(b) and (d), all the powers of T are bigger than 2. That is why the linewidths show a marked deviation from the standard anharmonic fit. To understand this unusual behavior, we note that strong electron correlation effects have been unveiled by ARPES in FeSe-based superconductors \cite{Yi}, and it is suggested these correlation effects enhance EPC \cite{Mandal,Gerber}. Pressure offers another means to increase electron carrier density and induce an emerging non-Fermi-liquid behavior \cite{Sun1,Okabe}, a sign of enhanced electron correlation in these systems. Moreover, temperature is also effective in tuning carrier density \cite{Sun,Sun2,Lei1} hence EPC \cite{Mandal,Coh,Gerber} and electron correlation effects in FeSe-based superconductors \cite{Sun1,Okabe,Mandal,Gerber,Yi}. Below we take on the challenging task of unraveling the highly convoluted physical phenomena and extract useful information to establish a clear picture for mechanism of superconductivity in FeSe-based compounds.

\begin{figure}[t]
	\includegraphics[angle=0,width=8cm]{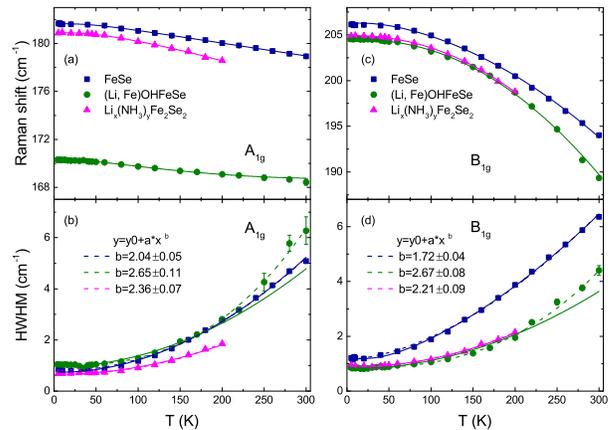}
	\caption{Temperature dependence of the $A_{1g}$ and $B_{1g}$ phonon modes in Li$_x$(NH$_3$)$_y$Fe$_2$Se$_2$, (Li,Fe)OHFeSe and FeSe. (a),(c) Phonon frequencies. (b),(d) Linewidths (HWHM). Solid curves correspond to standard anharmonic fitting \cite{Balkanski}. Dashed curves are the power-law fitting. }
	\label{fig3}
\end{figure}

In tuning $ T_{c} $ of FeSe-based superconductors by pressure \cite{Medvedev,Sun1}, doping \cite{Lei} and intercalation \cite{Guo,Ying,Burrard-Lucas,Scheidt,Lu,Krzton-Maziopa,Zhang,Sun2}, two factors are considered possibly playing deciding roles. The first is crystal structural features, such as anion height, and the second is carrier concentration. These two factors vary simultaneously, making it difficult to assess their individual influence on superconductivity. Raman scattering offers a powerful tool to probe and distinguish pertinent mechanisms since the measured phonon modes are sensitive to even the slightest structural changes. We note that the pressure coefficients of the $A_{1g}$ and $B_{1g}$ modes in FeSe are 4.71
$cm^{-1}$/GPa  and 5.95 $cm^{-1}$/GPa ,  respectively \cite{Massat}, the Gr\"{u}neisen parameters $\gamma$ are 1.04 and 0.9 for $A_{1g}$ and
$B_{1g}$ modes respectively \cite{Gnezdilov,Ksenofontov}, and the directional linear compressibility of the crystal, $K_{a}, K_{b}$ and $K_{c}$, are 0.63\%, 0.69\%, 1.76\% /Gpa \cite{Millican}. Because the lattice parameters \textit{a} and \textit{b} are basically unchanged and Se height is a half of the height of FeSe layer, the Gr\"{u}neisen law, $ \gamma=\frac{\mathrm{d}\ln\omega}{\mathrm{d}\ln V} $, allows an estimate of the relative rate of change of Se height from adjacent Fe layers, $h$, which is 0.48 times the relative frequency change of the $A_{1g}$ mode and 0.56 times that of the $B_{1g}$ mode. The above information also allows a convenient estimate of the effective pressure on the FeSe layer and its effect through the change of the phonon frequency from Raman scattering measurements. Since the $A_{1g}$ mode is susceptible to the influence of the charged intercalated layers and thus not a good probe for the intrinsic lattice dynamics of the FeSe layers, we take the $B_{1g}$ mode to estimate the effective pressure and the relative Se height. Results are summarized in Table I.

\begin{table}[!htbp]
	\centering
	\caption{Effective pressures and relative Se heights in Li$_x$(NH$_3$)$_y$Fe$_2$Se$_2$, (Li,Fe)OHFeSe and (K,Fe)Fe$_{2}$Se$_{2}$ derived from the frequencies of the $B_{1g}$ Raman mode. The effective pressures, $^aP$s and $^bP$s are estimated using the pressure coefficient 5.95 $ cm^{-1}$/GPa  \cite{Massat} and linear compressibility 1.76\% /Gpa \cite{Millican} through the Gr\"{u}neisen law. The relative Se heights $\Delta h=h-h_{Se}$ are estimated by the Gr\"{u}neisen law. All the values of $^bP/^aP$ are about 1.8 in the three samples. This indicates that the difference between $^aP$ and $^bP$ stems from the different parameters we used in the two methods, where the pressure coefficient of 5.95 $ cm^{-1}$/GPa  \cite{Massat} is an experimental value and the linear compressibility of 1.76\% /Gpa \cite{Millican} is a theoretical one.}\label{tab:1}
	

		\begin{tabular}{|c|c|c|c|}
		\hline
		     &(Li,Fe)OHFeSe&Li$_x$(NH$_3$)$_y$Fe$_2$Se$_2$&(K, Fe)Fe$_{2}$Se$_{2}$\\
		\hline
		$^aP$ (GPa)&-0.29&-0.23&2.80\\
		\hline
	    $^bP$ (GPa)&-0.54&-0.42&5.11\\
			\hline
		$\Delta h$ (pm) &0.70& 0.55& -6.68\\
			\hline
	\end{tabular}
\end{table}

In Fig. 4(a) we show the $B_{1g}$ frequencies and $T_{c}$ of the FeSe-based superconductors. The effective pressures relative to pristine FeSe are -0.23 and -0.29 $GPa$ for Li$_x$(NH$_3$)$_y$Fe$_2$Se$_2$ and (Li,Fe)OHFeSe, respectively. These effective pressures are too small to cause any significant increase in $T_{c}$ to explain the observed change from 8 to 44 K because $T_c$ only increases less than 2 K under 1 GPa pressure in FeSe \cite{Okabe}.  Furthermore, the effective pressure in (K,Fe)Fe$_{2}$Se$_{2}$ is 2.8 GPa, much higher than that in Li$_x$(NH$_3$)$_y$Fe$_2$Se$_2$, but both systems have nearly identical $T_{c}$ around 44 K. This demonstrates that pressure effect is not a key factor in deciding superconductivity in the intercalated FeSe superconductors.

The relative height of Se, \textit{h}, has been proposed to be a controlling parameter for superconductivity in FeSe-based superconductors.  Here we find that the Se heights in Li$_x$(NH$_3$)$_y$Fe$_2$Se$_2$ and (Li,Fe)OHFeSe are 0.38\% and 0.48\% higher with respect to \textit{h} in FeSe, corresponding to values less than 1 pm, and the difference of \textit{h} between the two intercalated superconductors is only about 0.15 pm. Meanwhile, \textit{h} in (K,Fe)Fe$_{2}$Se$_{2}$ is higher by 5\% (7 pm), much larger than that in Li$_x$(NH$_3$)$_y$Fe$_2$Se$_2$ , yet they have the same $T_{c}$ value. This result rules out Se height \textit{h} as the key factor for superconductivity in intercalated FeSe superconductors.

\begin{figure}[t]
	\includegraphics[angle=0,width=8cm]{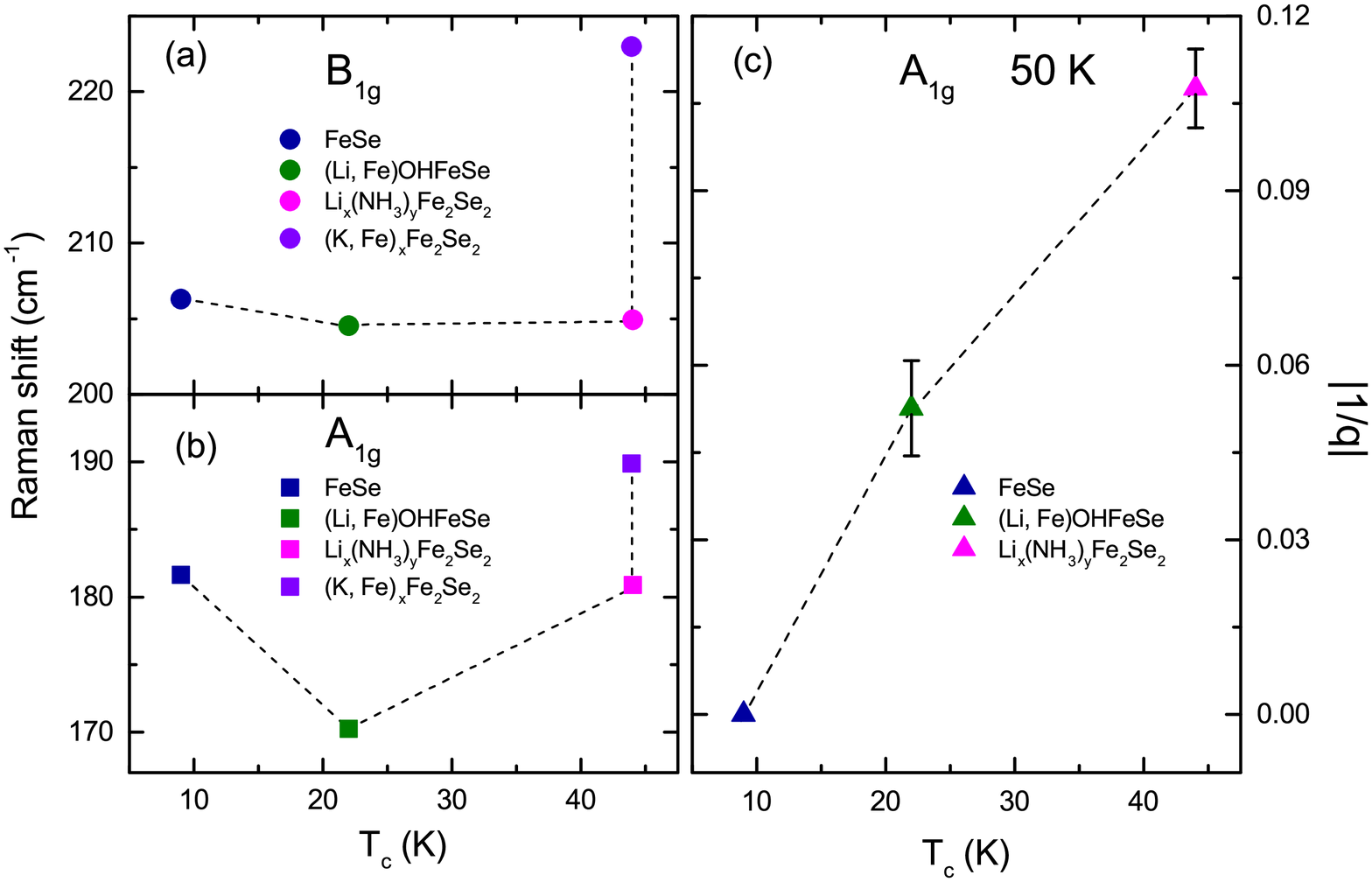}
	\caption{(a),(b) The relations between $T_{c}$ and saturation frequency of the $B_{1g}$ (Fe) and $A_{1g}$ (Se) modes. (c) The relation between $T_{c}$ and asymmetric parameter $|1/q|$ of the $A_{1g}$ (Se) mode at 50 K. The data on $|1/q|$ in (c) can be found in Supplemental Material, Fig. S2 \cite{SM}. }
	\label{fig4}
\end{figure}

The $A_{1g}$ mode actually offers an important clue for resolving the present challenge. The asymmetric line shape of the $A_{1g}$ mode \cite{Klein} stems from the strong coupling of the crystal lattice with the electrons near the Fermi surface, which has been studied theoretically \cite{Mandal,Coh} and experimentally \cite{Gerber}. The behaviors of the $A_{1g}$ asymmetry provide crucial information on electrons near the Fermi surface. Specifically, the asymmetry parameter $ |1/q| $, which determines the line shape \cite{Klein}, is given by $ |1/q|  \approx (T_e/T_p)\pi V\rho(E)$, where $ T_{p} $ and $ T_{e} $ are the scattering amplitudes for the decoupled phonons and the electronic continuum, respectively, \textit{V} is the matrix element for the interaction between the discrete excitation and the continuum, and $ \rho(E) $ is the joint density of states of the occupied and empty states near the Fermi surface. In the present case, $(T_e/T_p)V$ is almost the same for different systems as they have similar FeSe layered structures, and therefore any variation in $|1/q|$ can only come from $\rho(E)$. We plot the asymmetry parameters $|1/q|$ at 50 K versus $T_{c}$ in Fig. 4(c). Results show that $T_{c}$ exhibits a positive correlation with $ |1/q| $ or joint density of states $\rho(E)$, and this establishes a key connection between $T_{c}$ and the condition of electronic occupancy that can be tuned by effective carrier doping or applied pressure. This view is supported by the fact that the $T_{c}s$ of FeSe and (Li,Fe)OHFeSe are significantly enhanced by liquid gating from 8 to 48 K \cite{Lei} and from 24 to 43 K \cite{Lei1}, respectively. ARPES experiments also suggest elevated Fermi surfaces in FeSe/STO, (Li$_{0.8}$Fe$_{0.2}$)OHFeSe and Rb$_{x}$Fe$_{2-y}$Se$_{2}$ \cite{Niu}. In fact, all the existing methods for tuning superconductivity in FeSe-based superconductors \cite{Fang,Medvedev,Lei,Guo,Ying,Burrard-Lucas,Scheidt,Lu,Krzton-Maziopa,Zhang,Wang,Ge,Sun,Sun1,Lei1} adjust the carrier density and characters of electrons near the Fermi surface. The present findings identify the effective carrier doping, rather than the Se height \textit{h}, as the key deciding factor for modulating superconductivity in intercalated FeSe superconductors.

In summary, we have performed a systematic Raman study of pristine FeSe superconductor and its intercalated compounds Li$_x$(NH$_3$)$_y$Fe$_2$Se$_2$, and (Li,Fe)OHFeSe. By comparing to FeSe and K$_{0.8}$Fe$_{1.6}$Se$_{2}$, we find that the FeSe layers remain intact in the intercalated superconductors, which is further confirmed by the temperature dependence of the phonon frequencies and linewidths. The temperature dependence of the phonon linewidths deviates from standard anharmonic effects but can be well fitted by power laws. This behavior is attributed to EPC, which is evidenced by the Fano asymmetry of the $A_{1g}$ mode in Li$_x$(NH$_3$)$_y$Fe$_2$Se$_2$ and (Li,Fe)OHFeSe. These results represent the first observation of EPC in FeSe-based superconductors by Raman scattering.  Most importantly, we find that Tc is little influenced by the frequency of the $B_{1g}$ mode that is a sensitive indicator of the crystal structure change in the FeSe layer, but instead has a positive correlation with the $A_{1g}$ asymmetry parameter $|1/q|$ that depends on the joint density of states $\rho(E)$.  This observation demonstrates that effective carrier doping, rather than the Se height \textit{h} or other crystal structure parameters, is the key deciding factor for superconductivity in intercalated FeSe superconductors. The present approach may also offer an effective avenue to explore mechanisms of other novel superconductors.

This work was supported by the Ministry of Science and Technology of China (2016YFA0300504 and 2017YFA0302904) and the NSF of China (11604383, 11774419 and 11474357).

\end{document}